\documentclass[aps,12pts,pre,amsmath,amssymb,preprint,showkeys,showpacs]{revtex4-1}
\usepackage{graphicx}
\usepackage{bm}
\usepackage{mathptmx}
\usepackage{epstopdf}
\usepackage{booktabs}
\usepackage{multirow}
\usepackage{hhline}
\usepackage{dcolumn}
\usepackage{subfigure}
\usepackage{amsmath}
\usepackage{float}
\usepackage{chemformula}
\usepackage[version=3]{mhchem}
\usepackage{chemfig}
\usepackage{soul}
 
\begin{document}
\title
  {Scaling relation for the adsorption energies at bimetallic magnetic surfaces: Role of surface magnetic moment and work-function}

\author{Swetarekha Ram, Seung-Cheol Lee$^*$, and Satadeep Bhattacharjee}
\email{seungcheol.lee@ikst.res.in}
\email{satadeep.bhattacharjee@ikst.res.in}
\affiliation{Indo-Korea Science and Technology Center (IKST), Bangalore 560065, India}
\begin{abstract}
 
 The scaling relationships between the adsorption energies of different reaction intermediates have a tremendous effect in the field of surface science, particularly in predicting new catalytic materials. In the last few decades, these scaling laws have been extensively studied and interpreted by a number of research groups which makes them almost universally accepted. In this work, we report the breakdown of the standard scaling law in bimetallic transition metal (TM) magnetic surfaces for O and OH adsorbates, where adsorption energies are estimated using density functional theory (DFT). We propose that the scaling relationships do not necessarily rely solely on the adsorbates, they can also be strongly dependent on the surface properties.

\end{abstract}

\maketitle

The industrial scale synthesis of chemical compounds demands the use of catalysts as it facilitates the reaction by reducing energy barriers, without affecting chemical equilibrium\cite{spencer1983catalysis}. Transition-metal (TM) catalysts at their low dimensionality have gained incredible interest in recent years due to their unique chemical and physical stability \cite{wei2018heterostructured, mahmood2018electrocatalysts,vogiatzis2018computational}. The quantification of the adsorption of gas molecules on the surface of the substrate
is largely achieved by estimating their adsorption energy.  Prior to the development of a simple formulation of scaling relations \cite{abild2007scaling}, several studies have been reported similar concepts. The adsorption energy of hydrocarbons on late transition metals was predicted from the corresponding number of bonds to the surface \cite{kua2000thermochemistry}. The energy of adsorption of X atoms (X = C, S, O) on transition metal M, estimated by the energy of formation of the bulk MX species \cite{toulhoat2003kinetic}.  Examination of water electrolysis disclosed the linear correlation of O, OH, OOH on metal  \cite{rossmeisl2005electrolysis} and oxide  \cite{rossmeisl2007electrolysis} surfaces.  Apart from these, linear correlations between vibrational frequencies of adsorbates on TM surfaces were explained by  L. Joshua \textit{et. al} \cite{lansford2017scaling}. However, a thorough analysis of this
scaling revealed its limitations \cite{wang2017breaking, gani2018understanding, calle2017covalence, khorshidi2018strain, hong2016doped,  doi:10.1021/acscatal.6b03403}. Most of the TMs surface that explored
in the literature are non-magnetic in nature
and the role of magnetization of the TMs
surface to the adsorption energy and hence
towards the scaling laws are relatively unexplored. We calculated the adsorption energies of O and OH on a collection of magnetic bimetallic surfaces. We scrutinize the scaling relations in the estimation of their
adsorption energy. Our analysis involves the inclusion of surface magnetic moment and work function to the scaling laws which were earlier neglected. The other motivation behind selecting these particular bimetallic substrates is their novel tunable catalytic properties \cite{villa2015new, alshammari2016bimetallic, kitchin2004modification} as the interaction between two metal influence the electronic properties thereupon, its chemical properties \cite{kitchin2004role}. For comparison and credibility of our claim, we also calculated the adsorption energy of O and OH molecules for monoatomic TM surface and checked their affinity towards the standard scaling relations \cite{abild2007scaling,studt2008identification}.

   
 Spin-polarized density functional theory (DFT) calculations were performed using generalized gradient approximation(GGA) in the form of Perdew-Burke-Ernzerhof (PBE) for the exchange-correlation potentials \cite{perdew1996rationale, perdew1996phys}. The projector-augmented wave (PAW) method implemented in the Vienna Ab Initio Simulation Package (VASP) \cite{kresse1996efficient, kresse1996efficiency} was used for all calculations. The kinetic energy cutoff of 500 eV for a plane-wave basis set and $1\times10^{-6}$ eV for the self-consistent convergence were adopted. The optimized
lattice parameters and $k$-points were used
for the construction of (111) slab in our simulations. The vacuum region between the slabs normal to the
surface kept to be 20 $\AA$  to minimize the interaction between the periodic image. In the estimation of self-consistent $12\times 12\times 1$ Monkhorst-Pack grids \cite{monkhorst1976special} for Brillouin zone sampling was used in the surface. The gaseous O and OH molecule was optimized in a $10\times 10 \times 10$ $\AA$ cubic cell with a single $k$-point. The adsorbate coverage was chosen to 1/4 ML in all cases.
We also checked the reliability of our calculations
with larger supercells and with different coverage areas, where we find no significant difference. In atomic relaxation of
slab, the upper four layers, and adatoms were
allowed to relax until the force on each atom
was less than 0.06 eV/$\AA$, while the bottom
layers were kept fixed to mimic their bulk-like
behavior. 

 The adsorption energy ($ E_{ads} $) of O and OH on different types of system  can be expressed by  

\begin{equation}
E_{ads} = E_{slab+mol}- E_{slab} - E_{mol}
\end{equation}  

where $ E_{slab+mol} $ is the total energy of the adsorbate-covered slab, $ E_{slab}$ and $ E_{mol} $ represent the total energy for isolated clean slab and molecule, respectively.

A conjunction of adsorption energy ($E_{ads}$) between a partially hydrogenated adsorbate $AH_{X}$ ($E_{ads}^{AH_{X}}$) to its respective atomic adsorbate A ($E_{ads}^{A}$) across transition metal surfaces reveal from the discovery  of linear scaling relations, by N\o rskov and co-workers \cite{abild2007scaling}. i. e.

\begin{equation}
E_{ads}^{AH_{X}} = \gamma E_{ads}^{A} +\xi
\label{equ:scaling}
\end{equation}

\begin{figure}[]
  \centering
     \subfigure[]{\includegraphics[scale=0.6]{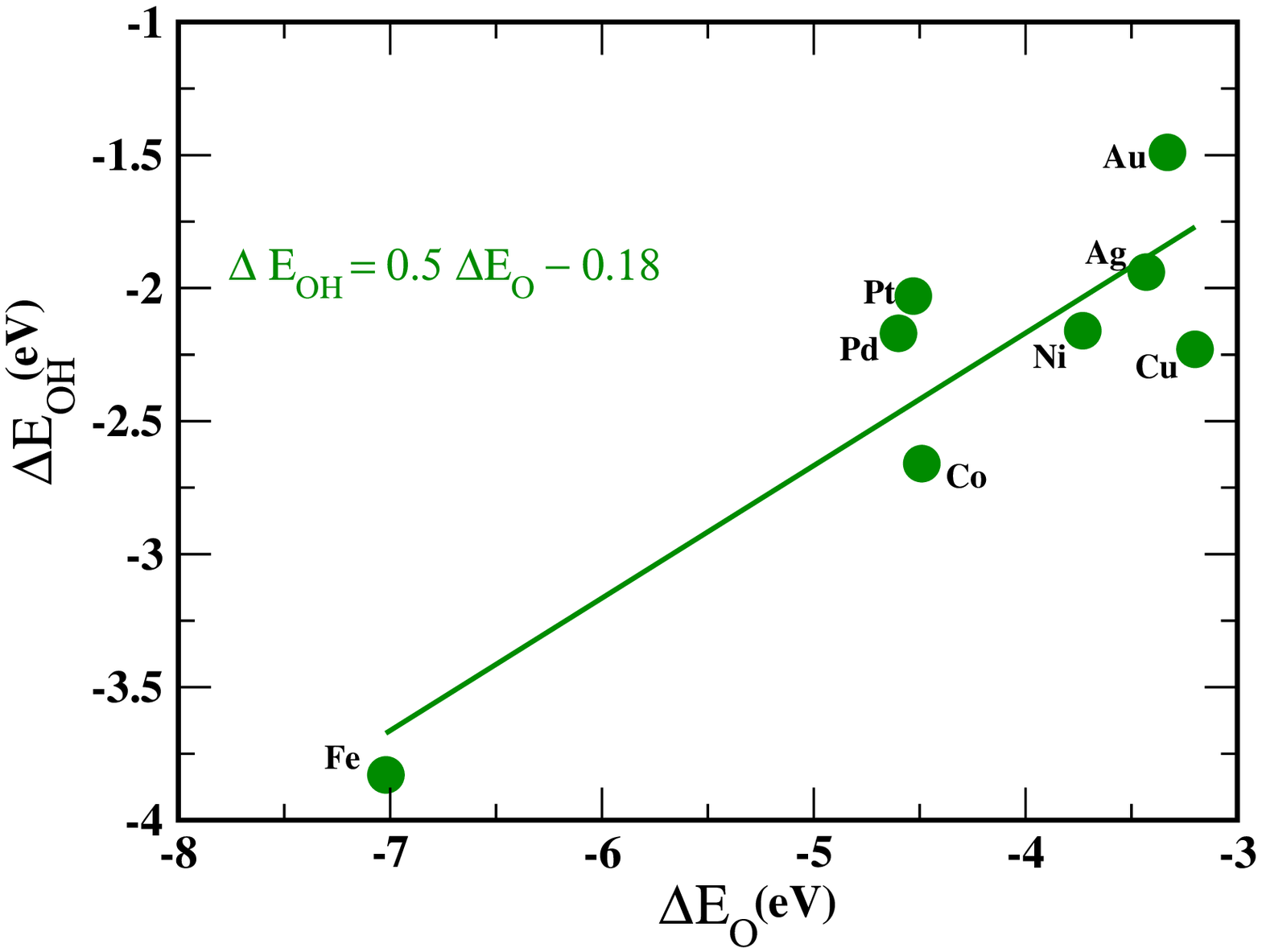}}    
    \subfigure[]{\includegraphics[scale=0.6]{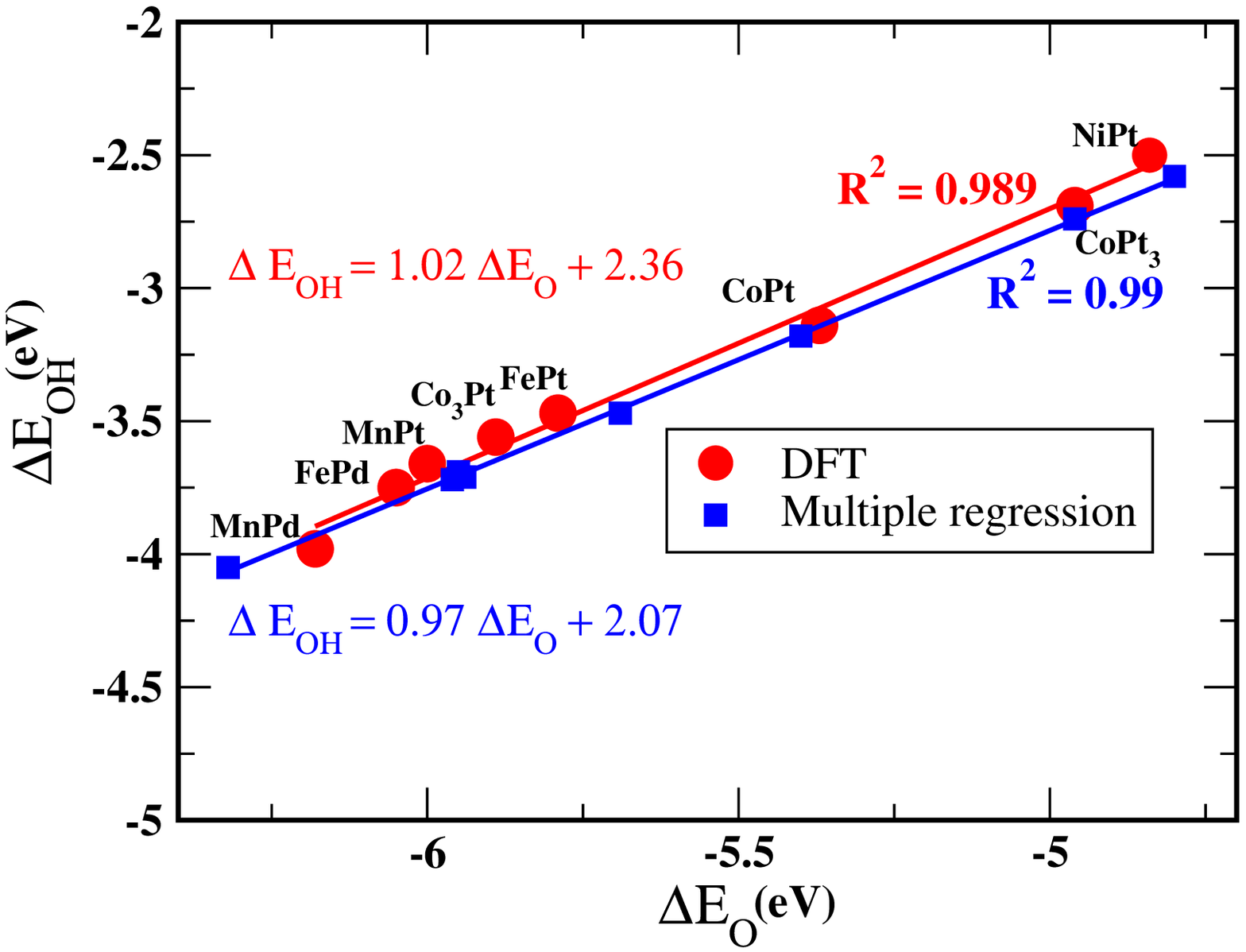}}
 \caption{Scaling relations for the adsorption energies of $ ^{*} $O and $ ^{*} $OH (a) for monoatomic tansition metal surface  (b) for TM magnetic bimetallic surface}
 \label{Fig:scalingOvsOH}
\end{figure}

\begin{figure}[]
  \centering
 
   \subfigure[]{\includegraphics[scale=0.7]{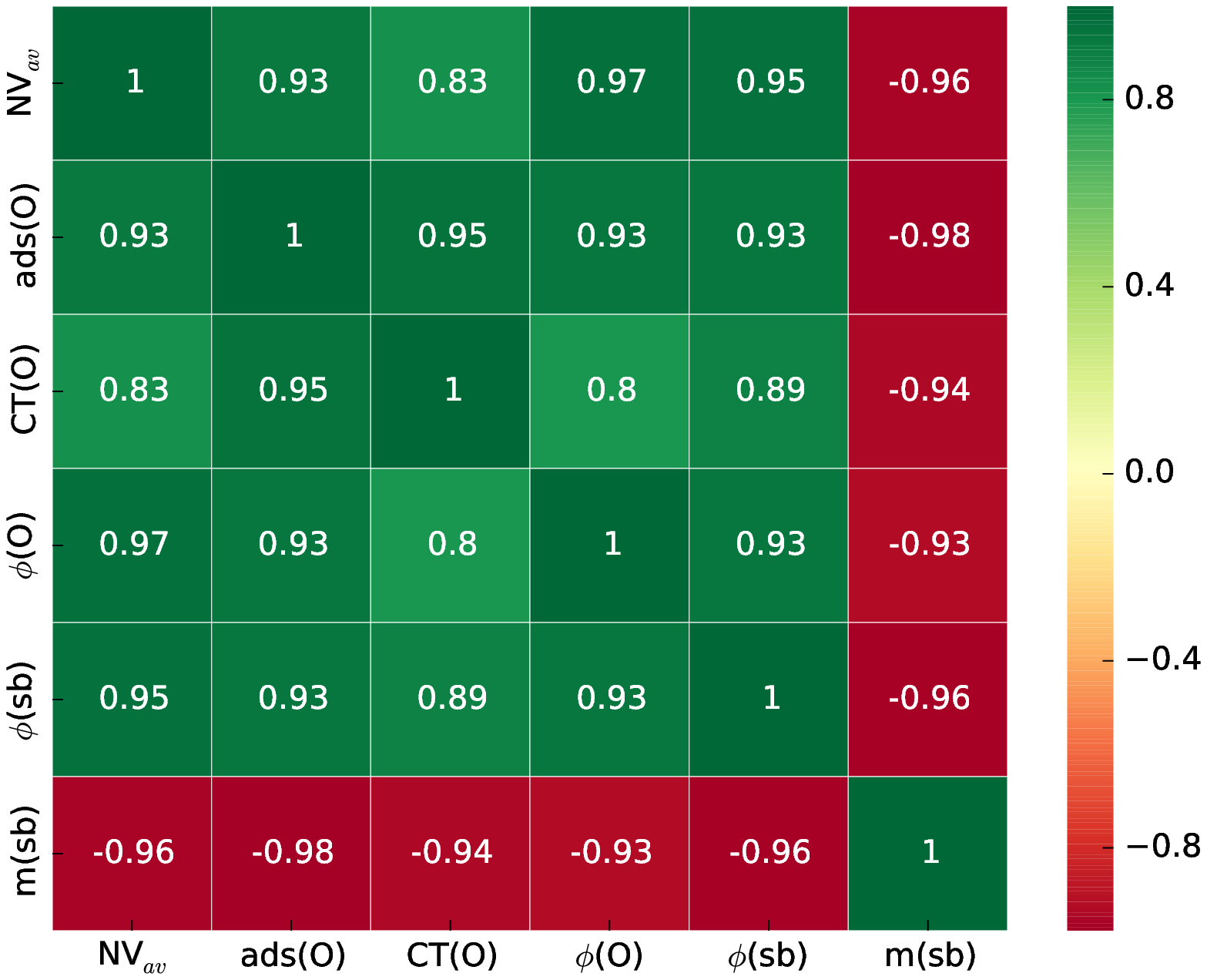}}
    \subfigure[]{\includegraphics[scale=0.7]{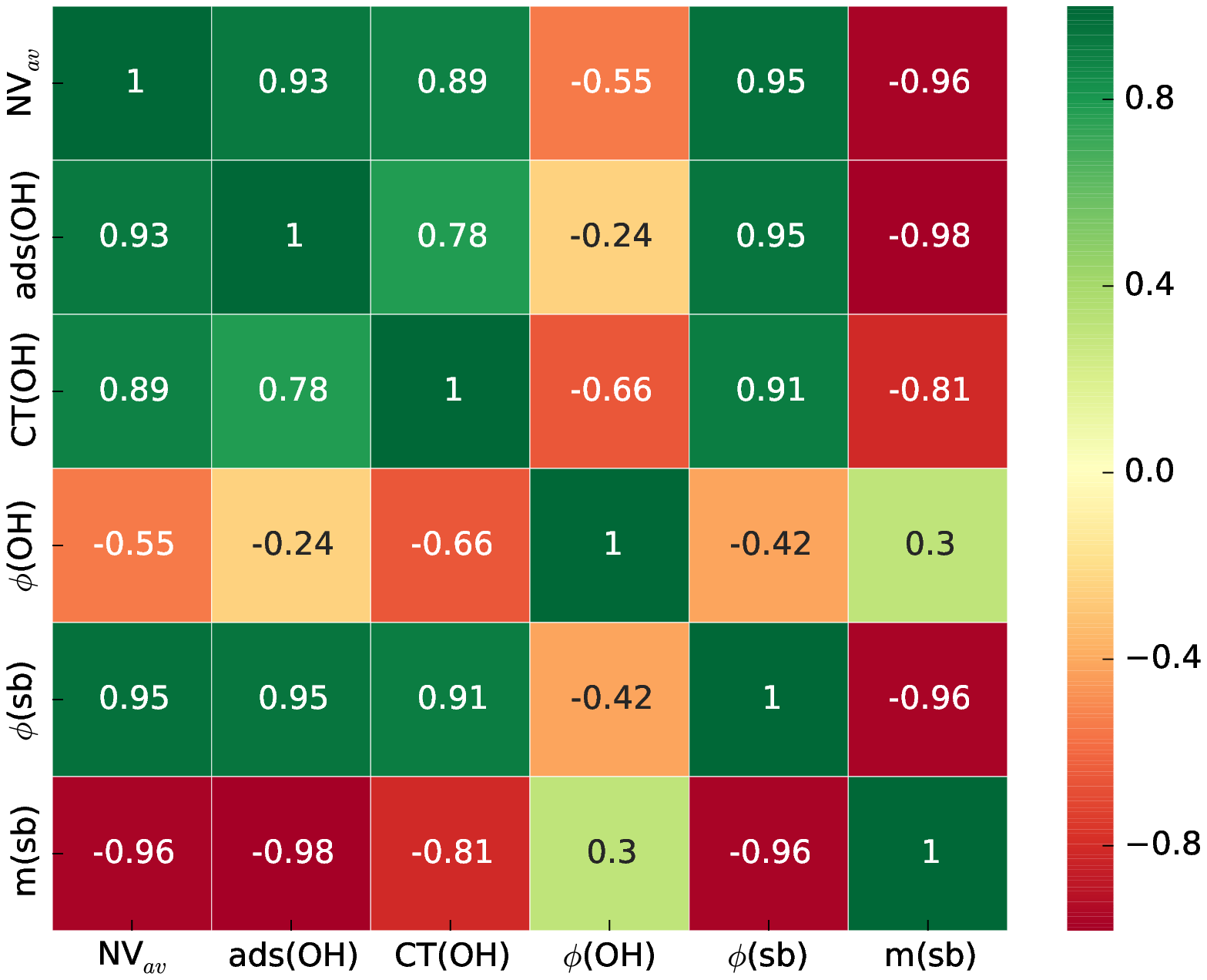}}
 \caption{Correlation matrix for different parameters (work function of clean slab, $ \phi (sb) $ and with O/OH adsorbed surface, $ \phi (O/OH) $, magnetic moment of slab (m(sb)), charge transfer (CT) with O/OH adsorbed surface, average valence elelctron ($ N_{av} $) of slab) associate with the (a) O adsorption and (b) OH adsorption.}
\label{fig:correlationmatrix}
\end{figure}

The slope, $\gamma$ depends on the valencies of A and $AH_{X}$ and has form: $\gamma = ({x_{max} - x})/{x_{max}} $ \cite{abild2007scaling}, where $x_{max} $ is the maximum number of hydrogen atoms that required to fulfill the octet rule for A and $x$ is the number of hydrogen atom present in $AH_{X}$. $\xi$ is the intercept, which depends upon slope \cite{calle2012physical} . 
Our calculated slope reaches the expected value of 0.5 for the monoatomic transition metal surfaces (Fig. \ref{Fig:scalingOvsOH} (a)) as predicted by N\o rskov group  \cite{abild2007scaling}.
But the scaling relation deviates from the expected in the case of TM bimetallic magnetic surfaces (see Fig. \ref{Fig:scalingOvsOH} (b) and the calculated values are in Table-I in supplementary information (SI).  The slope ($ \approx $ 1)  between the adsorption energies  of $ ^{*} $O and $ ^{*} $OH for the magnetic bimetallic surfaces in Fig. \ref{Fig:scalingOvsOH}(b) reflects the fact that both species deviate the bond formation between an O atom and the magnetic surface from the ideal case (2(O):O, 1(O): OH)  due to charge transfer. Hence it heightened a fact that how one can explicate the deviation in slope for $ ^{*} $O and $ ^{*} $OH adsorption.\\
\noindent In this study we propose the surface properties (magnetic moment, work function and number of average valence electrons of the surface) as a active parameters for the adsorption on bimetallic magnetic TM surface. If the adsorption energies of OH and O depend upon a set of surface variables, $\lbrace\omega_i\rbrace$, the $\Delta E_{OH}$ and  $\Delta E_{O} $ are interpreted  \cite{calle2012physical,calle2015introducing} as

\begin{equation}
\Delta E_{OH} = F(\lbrace\omega_i\rbrace) + \alpha_{0} \quad
 and \quad
\Delta E_{O} = G(\lbrace\omega_i\rbrace) + \beta_{0}
\label{equ:function}
\end{equation}

where $F$ and $ G$ are two functions of the set $\lbrace\omega_i\rbrace$ of the surface properties and $\alpha_{0} $ and $\beta_{0}$ depends on surface coordination number \cite{calle2015introducing,calle2014fast}, but in our case it is constant as we have considered adsorption on close packed surface. At the same time, the scaling relation described in equation-\ref{equ:scaling} must satisfy the equation-\ref{equ:scalingn} as below \cite{calle2012physical,calle2015introducing}, which contribute further physical and chemical perception.
\begin{equation}
F (\lbrace\omega_i\rbrace) = \gamma G (\lbrace\omega_i\rbrace)\
\label{equ:scalingn}
\end{equation}
It is important to find the parameters of the set $ \lbrace\omega_i\rbrace  $. 
To identify the important parameters,  $ \lbrace\omega_i\rbrace$, 
Pearson correlation coefficient matrices (Fig. \ref{fig:correlationmatrix}) are calculated from equation-\ref{equ:pearson}, which is a measure of the linear association between two variables $x$ and $y$, where $x, y \in  \lbrace\omega_i\rbrace $  and presented in Fig. \ref{fig:correlationmatrix} (see Table-II in SI).
\begin{equation}
  r (x,y) =
  \frac{ \sum_{i=1}^{n}(x_i-\bar{x})(y_i-\bar{y}) }{%
        \sqrt{\sum_{i=1}^{n}(x_i-\bar{x})^2}\sqrt{\sum_{i=1}^{n}(y_i-\bar{y})^2}}
\label{equ:pearson}        
\end{equation}

 Fig. \ref{fig:correlationmatrix} reveals the appropriate parameters which can scale between the two adsorbates are magnetic moment ($ m$), average number of valence electron ($ NV_{av} $) and work function ($ \phi $) of slab. It can be noticed that the magnetic moment of the slab has a negative correlation with all parameters except for $\phi (OH)$, the work function with OH adsorption (details discussed in the succeeding section).


To brighten this, we have analyzed the second level of scaling  i.e structure-energy relationship.  
The adsorption energies of O and OH depend upon $ \phi (sb) $,  $ N_{av} $ and $m$ of bimetallic magnetic surface (see Fig. 1 in SI), as a independent variables. If both $F$ and $G$ from equation-\ref{equ:scalingn} are negative or positive, then $ \gamma $ $>$ 0 (see Table-III in SI) and  here is how one can find the offset ($ \xi $) of equation-\ref{equ:scaling}, $ \xi = \alpha_{0} - \gamma \beta_{0} $, which depends on the slope ($ \gamma $) \cite{calle2012physical}. Again from their corresponding data matrix (Fig. \ref{fig:correlationmatrix}), a multiple regression was carried out to determine the individual contribution towards the adsorption process. 
%
%
Multiple regression analysis gives estimates for the coefficients (see equation-1 in SI) and the best result obtained in our case is: 

\begin{subequations}

\begin{equation}
    \Delta E_{OH} = -1.16 + 0.24\phi (sb) - 0.21 N_{av} - 1.06 m(sb) 
\end{equation}

\begin{equation}
\Delta E_{O} = -4.39 + 0.31\phi (sb) -0.15 N_{av} - 0.99 m(sb) 
\end{equation}
\label{equ:multi_result}
\end{subequations}
  
Significance of magnetic moment, $m$ towards the adsorption energy for O and OH is noticeable from the coefficient of 
 $m $ (-1.06:OH;  -0.99:O). 

At the same time our DFT calculated adsoption energy amenable with the result obtained from equation-\ref{equ:multi_result} and is clearly evident from Fig. \ref{Fig:scalingOvsOH}(b).   
From the above observation we conclude that the generalized scaling relations not only depend on the adsorbate, but also depend on the surface property. After further analysis to the mathematical nature of the functions $F$ and $G$ (see Table-III in SI), the existance of slope to be unity for  $ ^{*} $O and $ ^{*} $OH is cleared and the Equation-\ref{equ:scaling} can be expressed as 
\begin{equation}
 \Delta E_{OH} = \gamma (\phi, NV_{av}, m)  \Delta E_{O} +\xi \\
\label{equ:scaling2}
\end{equation} 
 instead of, 
 $\Delta E_{OH} = \gamma (x_{max}, x)  \Delta E_{O} +\xi $ as proposed originally by Abild-Peterson \textit{et. al} \cite{abild2007scaling}.
 
%

It is crucial to resolve and analyze the reason for this anomalous behavior of slope in the case of  bimetallic TM magnetic surfaces. 
     %
     Our estimated magnetic moment for the bimetallic TM magnetic surfaces, show an enhanced magnetic moment with respect to its individual component surfaces. For example, in FePt the Fe atom has average magnetic moment 3 $\mu_{B}$ and Pt has 0.37 $\mu_{B}$, clearly showing an enhanced moment with respect to their atomic bulk phase which is estimated to be 2.31 $\mu_{B}$ for Fe and  0.001 $ \mu_{B} $ for Pt.    
             This is due to the electron transfer from the minority spin of Fe atoms to Pt and is accompanied by gain in magnetic moment of Pt, while it is the majority spin of Fe-$d$ and Pt-$d$ states which implicates that its strongly hybridized with $p$ states of O (see overlap of orbitals only in majority spin channel in Fig. \ref{dos} (a)). This pushes the majority spin states of Fe to lower energy, resulting in the enhancement of their occupation and hence in the magnetic moment \cite{bhattacharjee2016improved, bhattacharjee2016nh3}. Interestingly the scenario is not seen for the adsorption of O and OH in the case of monoatomic transition metal surface for both magnetic and non-magnetic elements (see Fig. \ref{dos} (b, c)). The majority and the minority spin act in a similar way in the adsorption process, as we find the contribution at the Fermi level from both the spin channel is nearly the same (see Fig. \ref{dos}(b) ). At the same time, the overlap of O-$p$ states with Fe-$d$ is not observed in both spin channel in the case of monoatomic magnetic transition metal. Furthermore, Fig. \ref{dos} suggests that dissimilar orbital overlaps between adsorbate and the metal surface in the case of monoatomic TM surface and bimetallic TM surface cause a difference in scaling relation. 

\begin{figure*}[ht]
  \centering
  \subfigure[]{\includegraphics[scale=0.5]{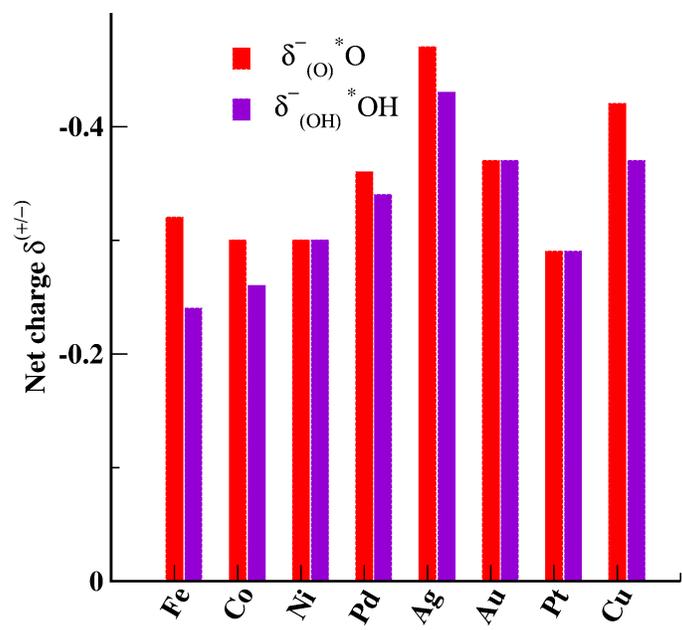}}
  \subfigure[]{\includegraphics[scale=0.5]{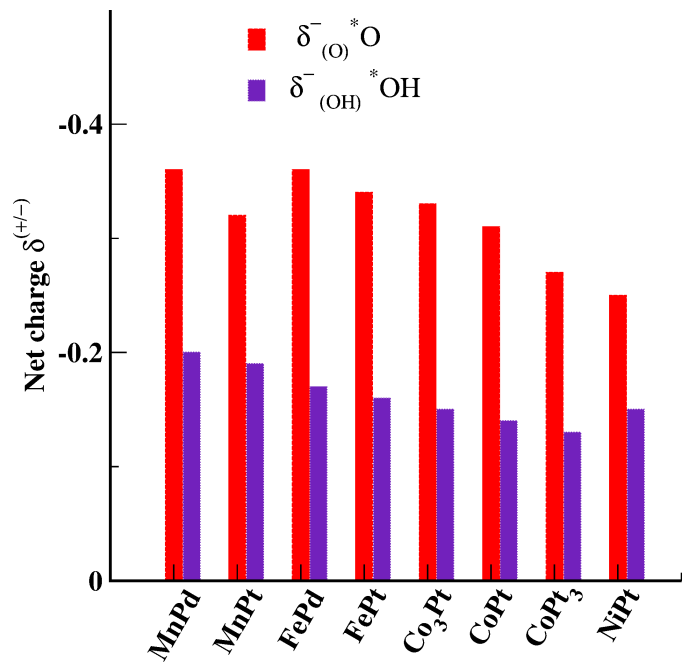}}
 \caption{Comparison of charge state (a) for monoatomic TM surface, (b) for bimetallic magnetic TM surface. 
 Net charge transfer for $ ^{*} $O and $ ^{*} $OH is approximately the same for a particular element in monoatomic TM surface, whereas it differs for magnetic bimetallic TM surface. }
 \label{fig:CT}
\end{figure*}

\begin{figure}[]
  \centering
  \subfigure[]{\includegraphics[scale=0.45]{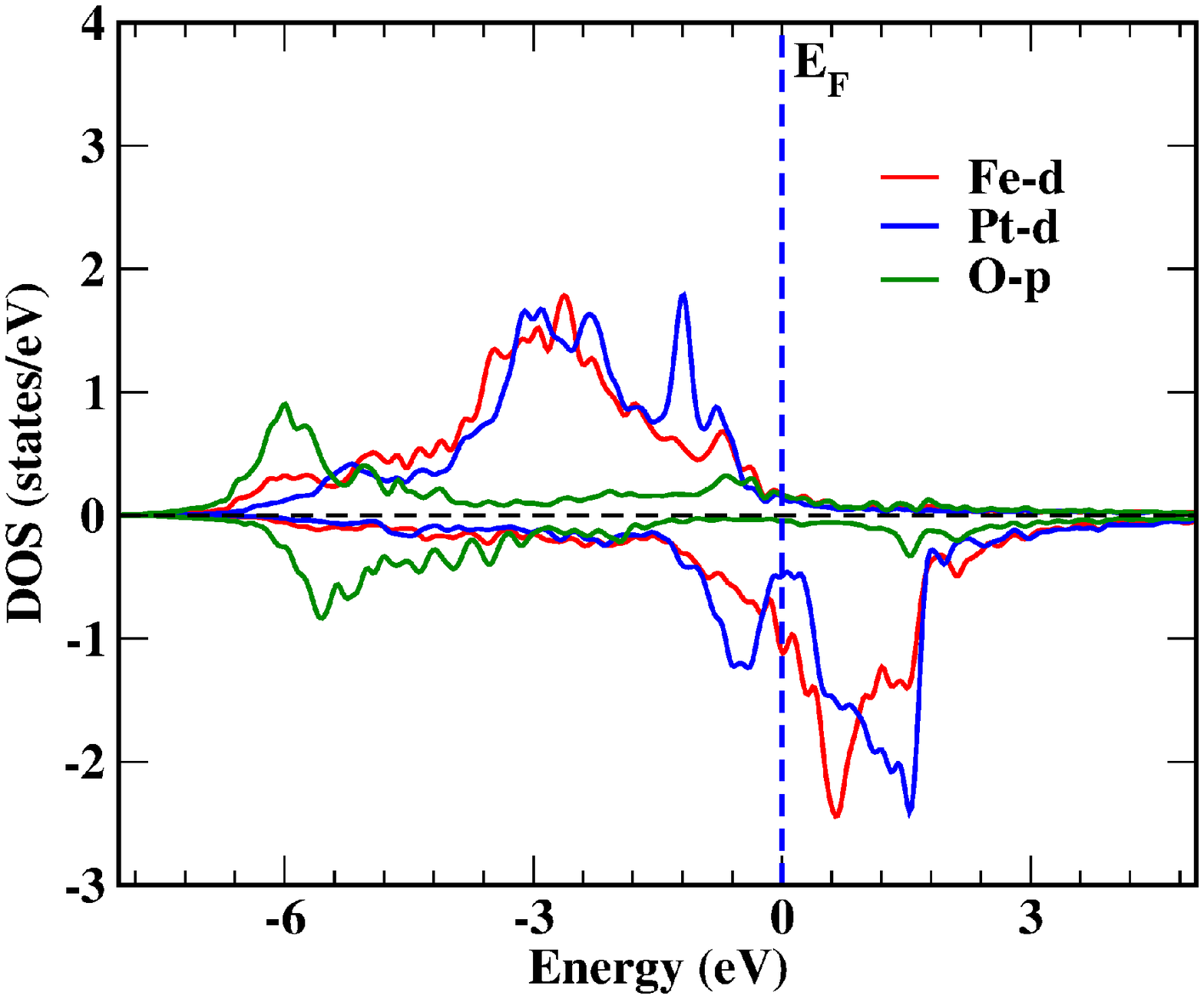}}
  \subfigure[]{\includegraphics[scale=0.45]{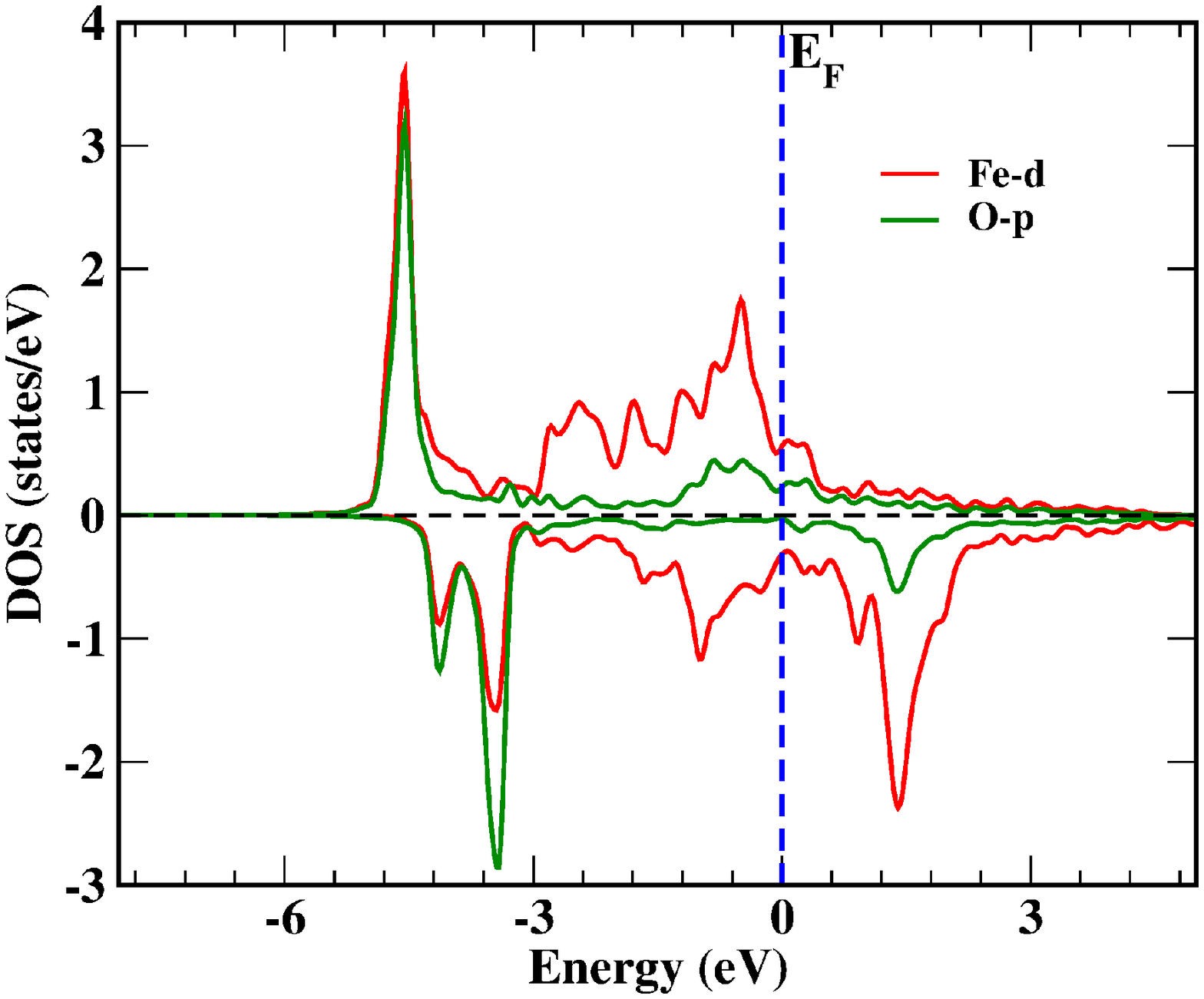}}
  \subfigure[]{\includegraphics[scale=0.45]{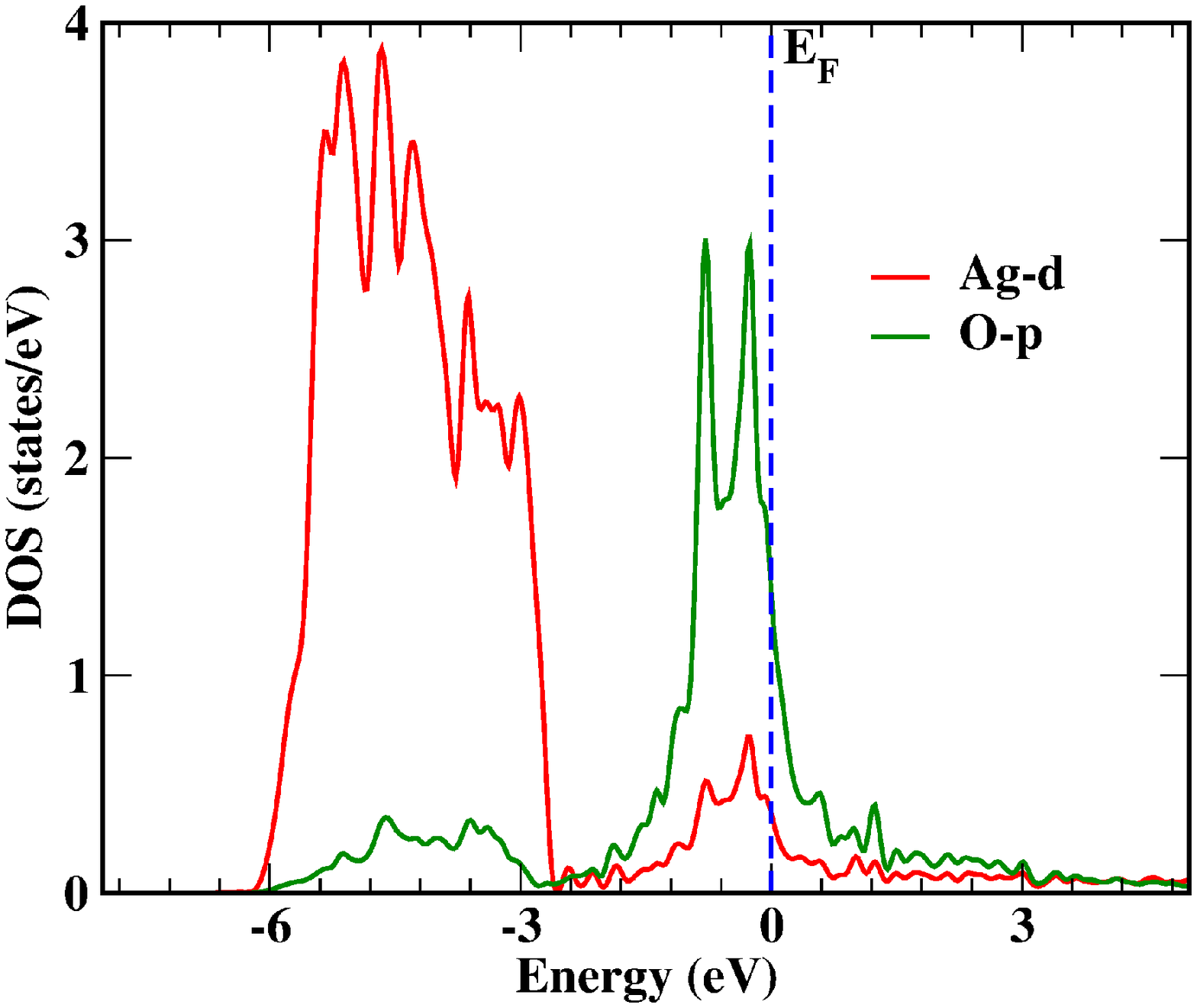}}
 \caption{DOS for (a) FePt with $ ^{*} $O adsorption,  (b) for Fe with  $ ^{*} $O  adsorption, (c) for Ag with  $ ^{*} $O  adsorption. The hybridization of Fe-$d$ with Pt-$d$ states is significant in majority spin only for FePt, whereas charge transfer is evident from the minority spin of FePt and in both spin chanel in Fe and for Ag also. It is the majority spin $d$-states which strongly hybridized with $p$-states of O-atoms in FePt.  The DOS profile remains almost similar as FePt for other bimetallic magnetic TM surface and for monoatomic TM, it is similar with Fe and Ag. }
  \label{dos}
\end{figure}

   O and OH are paramagnetic molecules, with a magnetic moment of 2 and 1 $\mu_{B}$ in the gas phase. When O and OH molecule is adsorbed on a bimetallic magnetic surface, the total magnetic moment on O and OH molecule is reduced (O:$\sim 0.1 \mu_{B}$; OH:$\sim 0.05 \mu_{B}$), with the majority of the moment located on the slab. As discussed by  Leenaerts et al. \cite{leenaerts2008paramagnetic},  the change in the magnetic moment of an adatom can be a useful indication of charge transfer.  To illustrate this, we have calculated the net charge using two different methods, as described below.
Density Derived Electrostatic and Chemical (DDEC6)~\cite{limas2016introducing, manz2010chemically} method and Bader charge analysis method developed by the Henkelman's group \cite{henkelman2006fast} have been carried out for the energetically most stable configuration to produce chemical states of atoms. The trend in charge transfer (CT) is noticed to be similar in both the methods, but the magnitude is different (see Table-IV in SI) as seen earlier \cite{iyemperumal2017activation}.
We find the excess negative charge on  O and OH approximately the same for a particular TM in monoatomic TM surface (see Fig. \ref{fig:CT}(a)), but in the case of bimetallic surfaces (see red bar in Fig. \ref{fig:CT}(b)) it is higher in O in comparison to OH.

\begin{figure}[]
  \centering
  \subfigure[]{\includegraphics[scale=0.20]{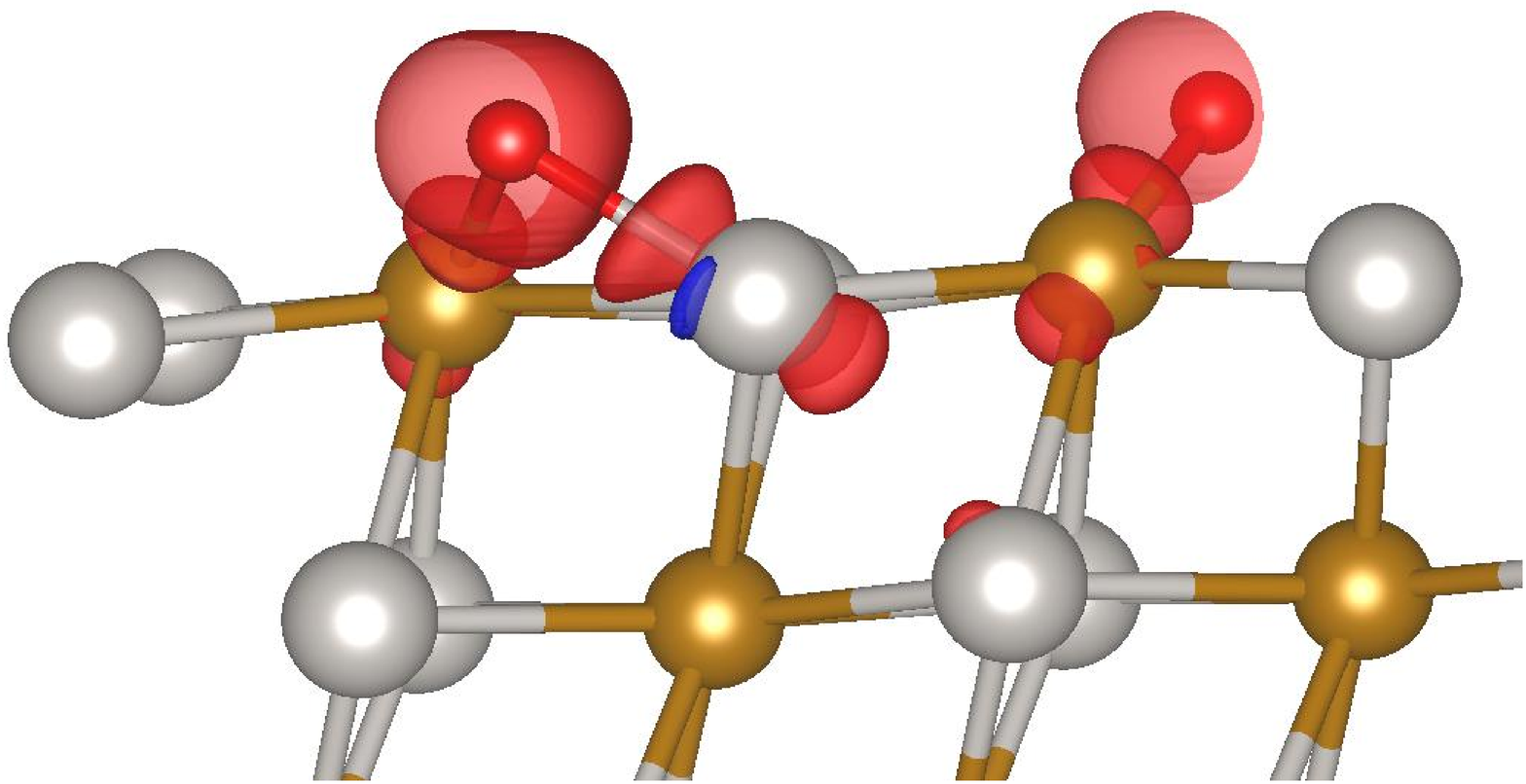}}
  \subfigure[]{\includegraphics[scale=0.20]{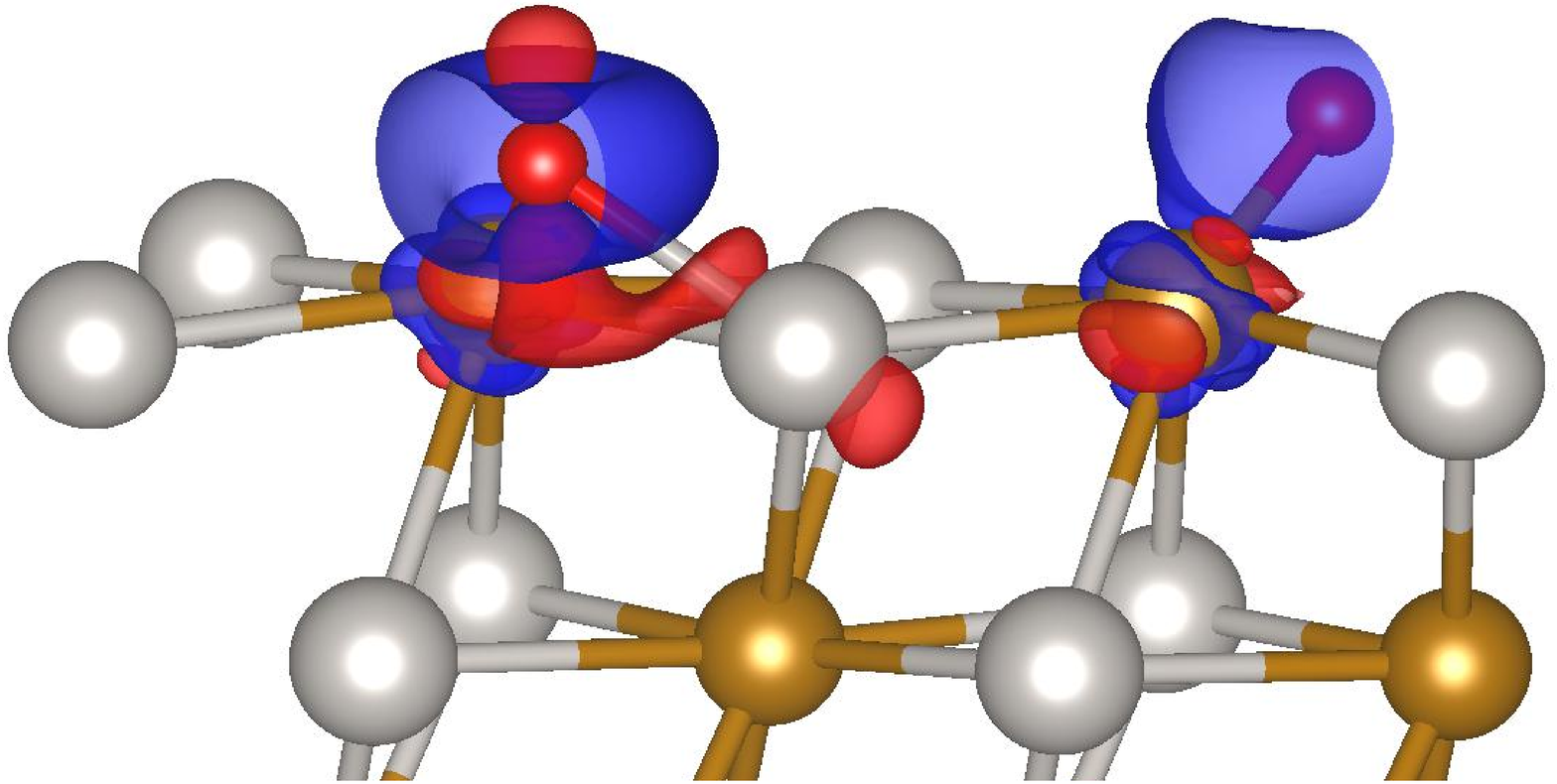}}
  \subfigure[]{\includegraphics[scale=0.20]{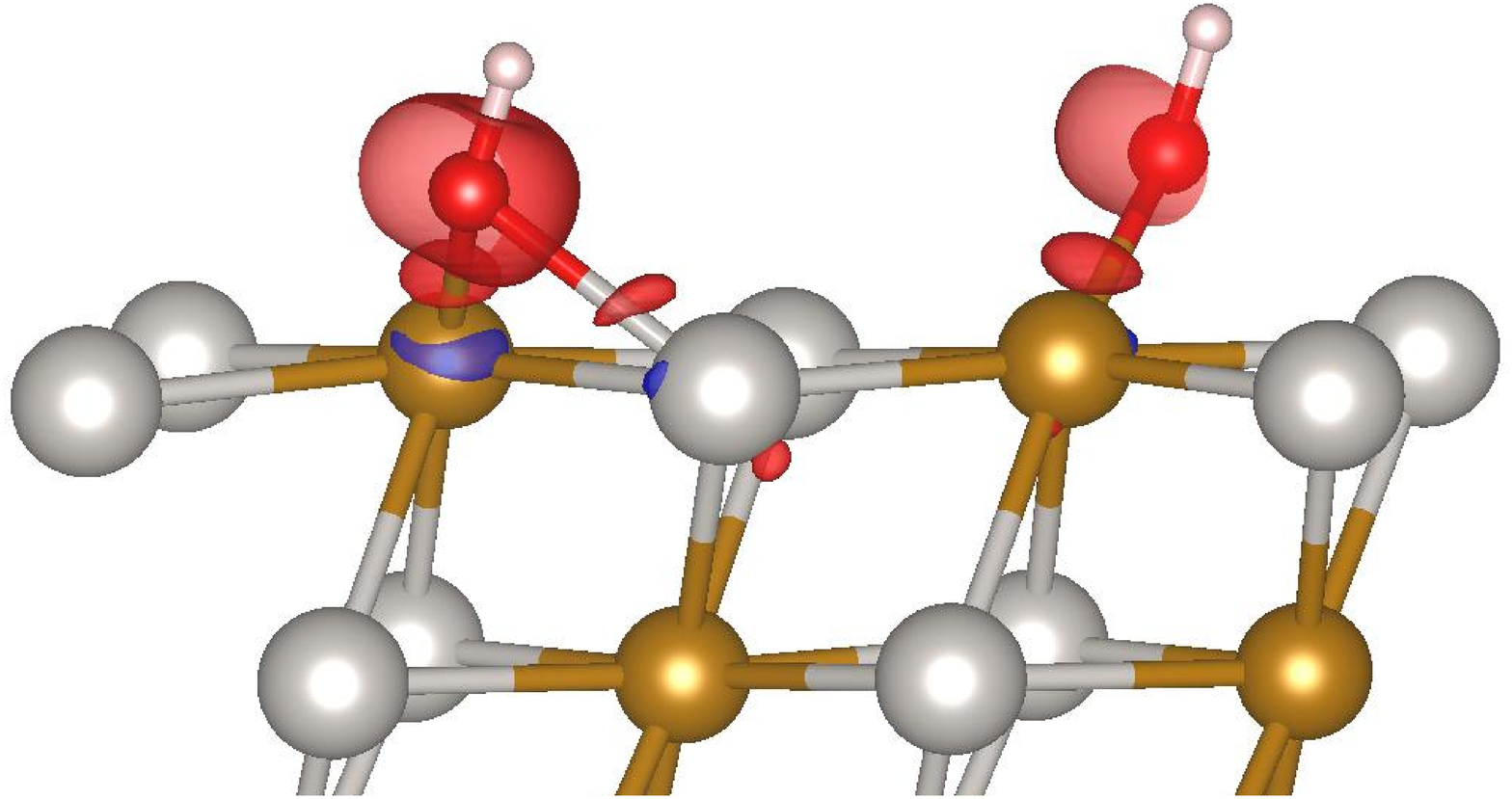}}
  \subfigure[]{\includegraphics[scale=0.20]{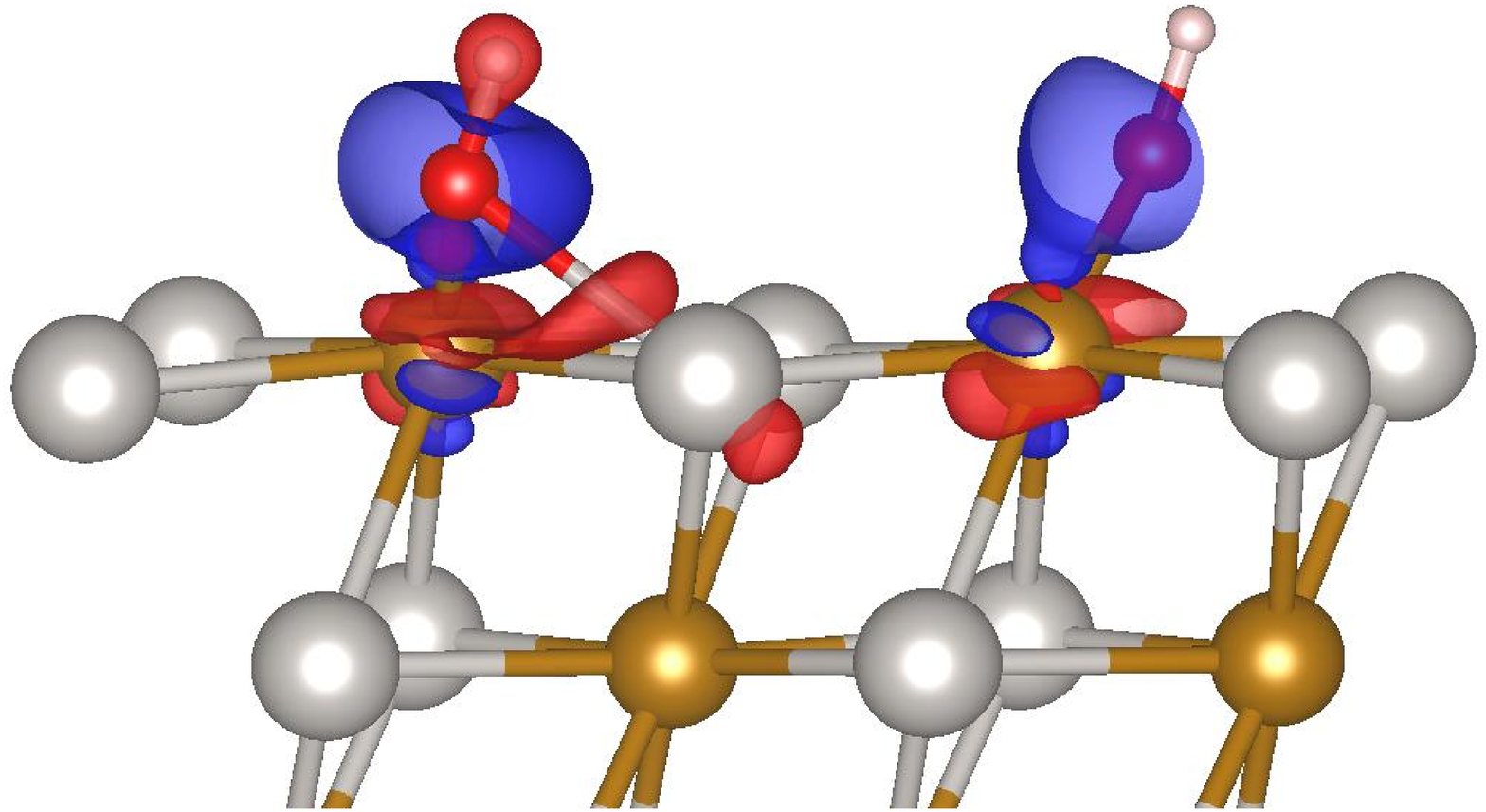}}
 \caption{Iso-surface of differential charge density for (a, c) majority spin of FePt with $ ^{*} $O and $ ^{*} $OH adsorption, respectively (b, d) minority spin of FePt with $ ^{*} $O and $ ^{*} $OH adsorption, respectively. The red and blue colors represent the electron accumulation and depletion, respectively. Charge density isosurface was set to 0.005 $e\AA^{-3}$. The sphere in golden, silver, red and white color represents Fe, Pt, O, H atoms respectively. The accumulated charge in $ ^{*} $O  is more abundant than that in $ ^{*} $OH, suggesting that the chemical bond for the $ ^{*} $O adsorption is more stable than that for $ ^{*} $OH adsorption. A similar observation is also noticed for remaining materials under study.}
 \label{fig:CDD}
\end{figure}

 The electrostatic interaction of O(OH)
on TM bimetallic magnetic surface is closely
  related to the charge transfer 
    ($E_{ads}$ is directly propertional to the CT ($\delta^{-}_{O/OH}$).  
  Our observations are seen to commensurate with the noticed trend in the case of  TM atoms on rutile TiO$_{2}$ (110) surface \cite{cai2013transition}.
To obtain a better insight into the charge redistribution after the adsorption processes, we find it instructive to plot (see Fig. \ref{fig:CDD} ) the spin-charge densities difference (SCDD) as

\begin{equation}
\Delta\rho_{\sigma}(r) = \rho_\sigma[slab+mol]- \rho_\sigma[slab] -\rho_\sigma[mol]
\end{equation}

Where $\sigma $ denotes majority and minority spin states.
To further understand the charge redistribution in the system, and to strengthen our result we compute the integral of the SCDD over the xy-plane (see Fig. 2 in SI). 
It  clearly validates that the charge transfer is considerably higher in $ ^{*} $O adsorption in comparison to $ ^{*} $OH. 
The adsorption energy ($E_{ads}$) is in general, a mixture of covalent and ionic contributions, as the band hybridization and electron transfer could occur simultaneously during the process~\cite{shen2017more}. The ionic term can be related to the work function ($\phi$)of the metal surface. As described above, the electron transfer occurs from Fe to Pt (Fig. \ref{dos}) in the bimetallic magnetic TM surface, again supports our calculated value of $\phi$. A decrease in $\phi$ for FePt clean surface ($\phi$ = 4.85 eV) in comparison to Pt clean surface ($\phi$ = 5.65 eV) is due such charge transfer from  Fe to Pt. The scenario is true for all bimetallic surfaces. Therefore, the work function plays an important role in chemisorption process~\cite{menamparambath2014large} and in scaling relationship.\\   
\noindent In conclusion, we report breakdown of the standard scaling relation in the estimation of adsorption energies of O and OH adsorbates. We suggest the cause for the breakdown of the law is due to previous biased consideration of the adsorbate properties to be the lone parameters in the estimation of adsorption energy. We showed that by including the surface magnetic moments and work-function of the substrate, the scaling relation can be recalibrated for bimetallic magnetic systems. Thus, in general, we provide an accurate scaling relation for estimating adsorption energy of similar adsorbate, O and OH on magnetic transition metal surfaces.

This work was supported by NRF grant funded by MSIP, Korea (No. 2009-0082471 and No.
2014R1A2A2A04003865), the Convergence Agenda Program (CAP) of the Korea Research Council of Fundamental Science and Technology (KRCF)and GKP (Global Knowledge Platform) project of the Ministry of Science, ICT and Future Planning.

\bibliography{ADS}


\newpage
\clearpage
\newpage
\clearpage
\newpage
 \clearpage
\newpage
    
\end{document}


\preprint{}

\title{Supplementary Material:\--- Scaling relation for the adsorption energies at bimetallic magnetic surface: Role of
surface magnetic moment and work-function}
\author{Swetarekha Ram, Seung-Cheol Lee$^*$, and Satadeep Bhattacharjee}
\email{seungcheol.lee@ikst.res.in}
\email{satadeep.bhattacharjee@ikst.res.in}
\affiliation{Indo-Korea Science and Technology Center (IKST), Bangalore 560065, India}


\date{\today}

\pacs{Valid PACS appear here}
\keywords{}
\maketitle


%
\section{Adsorption energy of studied systems}

The calculated adsorption energy for O and OH molecule in bimetallic magnetic transition metal (TM) surface and the mono-atomic TM surface in close-packed structure is presented in Table-I.

\begin{table}[!h]
\caption{Adsorption energy, $E_{ads}$ for stable site of bimetallic TM magnetic systems and mono-atomic TM in close packed surface, for $ ^{*} $O and $ ^{*} $OH adsorbate.}
\label{tab:my-table}
\begin{tabular}{|c|cc|c|cc|}
\hline
\multirow{2}{*}{Systems } & 
     \multicolumn{2}{c|}{E$_{ads}$ (eV)} &
     \multirow{2}{*}{Systems} & 
     \multicolumn{2}{c|}{E$_{ads}$ (eV)} \\     
             
      \cline{2-3}
      \cline{5-6}
      
    & $ ^{*} $O \hspace{0.5mm} & $ ^{*} $OH &       &$ ^{*} $O & $ ^{*} $OH       \\             \hline
MnPd		 & -6.08 \hspace{1mm}	& -3.98		& Fe	   & -7.02	\hspace{1mm}	& -3.83		\\
MnPt         & -6       & -3.66     & Pd       & -4.6       & -2.17    \\
FePd         & -6.05    & -3.75     & Pt       & -4.53      & -2.03    \\
FePt         & -5.79    & -3.47     & Co       & -4.49      & -2.66    \\
Co$_3$Pt     & -5.89    & -3.56     & Ag       & -3.73      & -2.16    \\
CoPt         & -5.37    & -3.14     & Ni       & -3.43      & -1.94     \\
CoPt$_3$     & -4.96    & -2.96     & Au       & -3.33      & -1.49     \\
NiPt         & -4.84    & -2.5      & Cu       & -3.2       & -2.23     \\ 
\hline             
\end{tabular}
\end{table}

\subsection{Parameters associated with scaling relation}
Pearson correlation coefficient matrices are calculated for the bimetallic magnetic TM surface, which is a measure of the linear association between two variables $x$ and $y$, where $x, y \in \lbrace\omega_i\rbrace$. The parameters of the set $ \lbrace\omega_i\rbrace $ is presented in Table-II.

\begin{table*}
\caption{Average valence electron ($NV_{av}$), work function $\phi$(sb) of the clean slab, as well as with the presence of adsorbates $\phi$(O/OH), magnetic moment of slab, m(sb), net charge transfer (CT) with the presence of O and OH molecule and the adsorption energy of O and OH adsorbate in magnetic bimetallic TM surface.    }
\label{tab:my-table}
\begin{tabular}{|l|l|l|l|l|l|l|l|l|l|}
\hline
Systems	& $NV_{av}$ & $\phi(sb)$ & m(sb) & $E_{ads}(O)$ & CT(O) & $\phi$(O) & $E_{ads}(OH)$ & CT(OH) & $\phi(OH)$ \\
\hline
MnPd	& 8.5    & 4.47    & 2.06  & -6.18     & -0.36 & 5.37   & -3.98      & -0.2   & 4.17    \\
MnPt	& 8.5    & 4.72    & 1.77  & -6        & -0.32 & 5.29   & -3.66      & -0.19  & 4.55    \\
FePd	& 9      & 4.72    & 1.7   & -6.05     & -0.36 & 5.42   & -3.75      & -0.17  & 3.26    \\
FePt	& 9      & 4.85    & 1.72  & -5.79     & -0.34 & 5.48   & -3.47      & -0.16  & 3.43    \\
Co$_3$Pt& 9.25   & 5.03    & 1.49  & -5.89     & -0.33 & 5.54   & -3.56      & -0.15  & 3.26    \\
CoPt	& 9.5    & 5.11    & 1.18  & -5.37     & -0.31 & 5.69   & -3.14      & -0.14  & 3.52    \\
CoPt$_3$& 9.75   & 5.33    & 0.77  & -4.96     & -0.27 & 5.78   & -2.69      & -0.13  & 3.65    \\
NiPt	& 10     & 5.33    & 0.57  & -4.84     & -0.25 & 5.75   & -2.5       & -0.15  & 3.62      \\
\hline  
\end{tabular}\\

\end{table*}


\begin{figure}[h]
  \centering
  \subfigure[]{\includegraphics[scale=0.3]{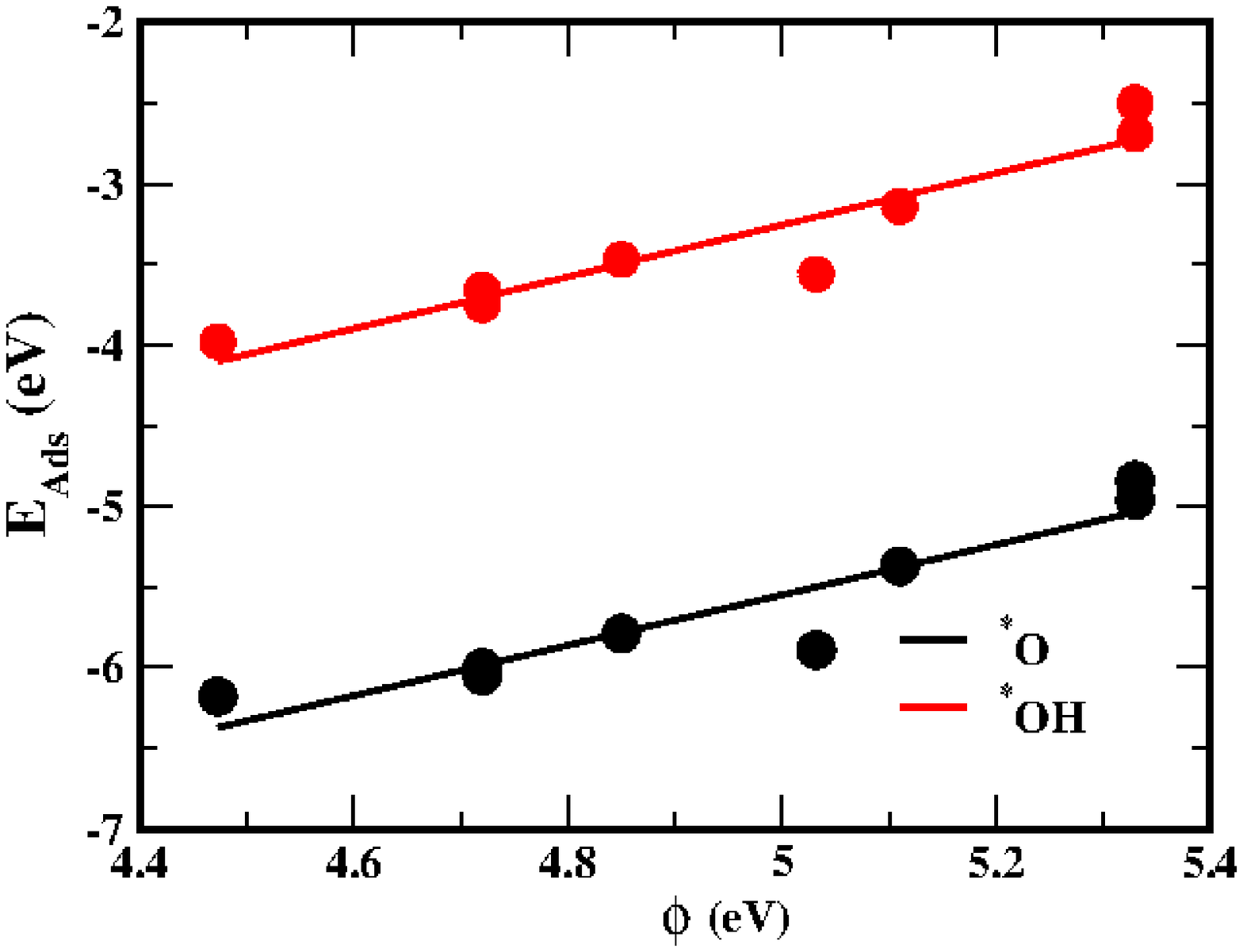}}   
   \subfigure[]{\includegraphics[scale=0.3]{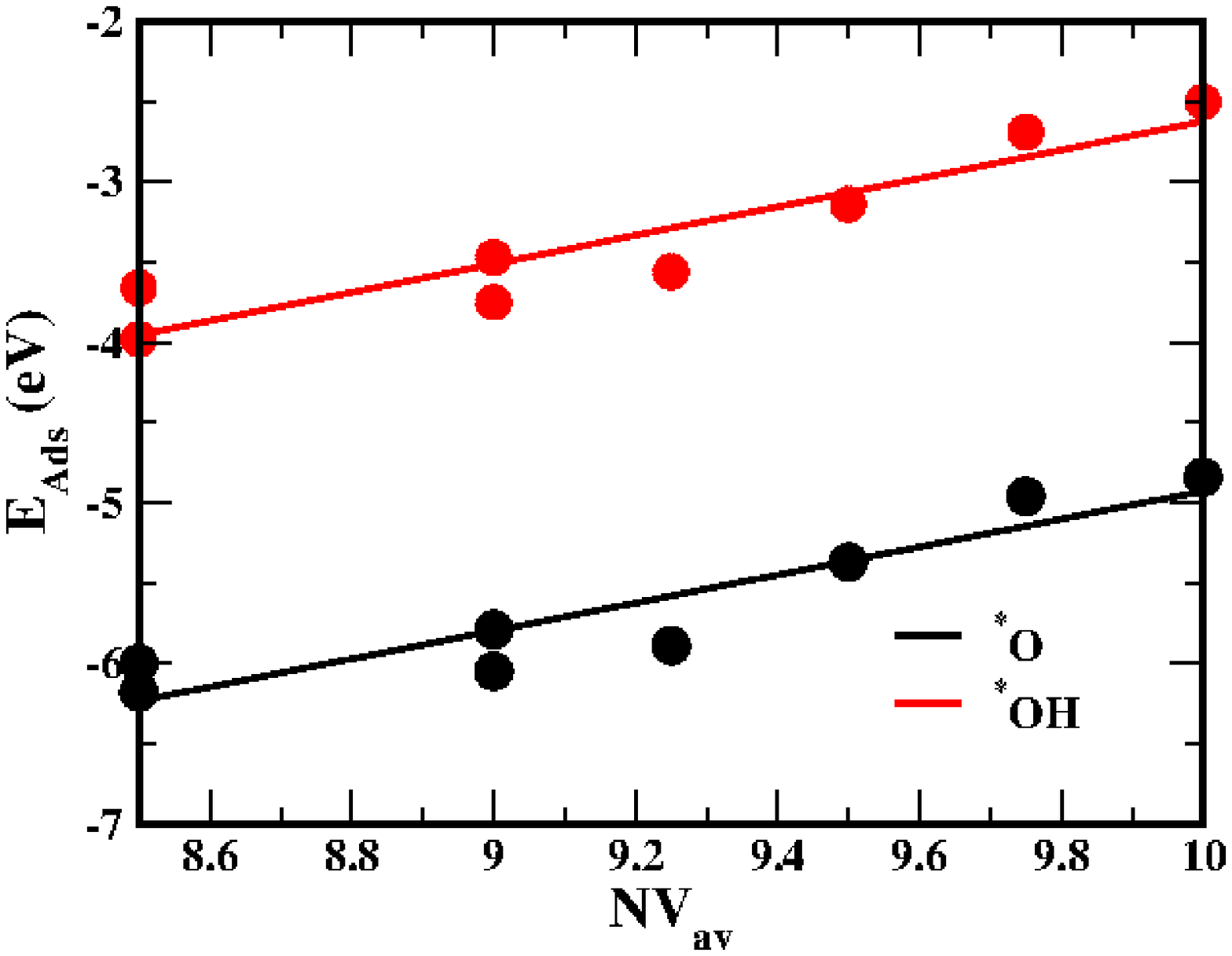}}
     \subfigure[]{\includegraphics[scale=0.3]{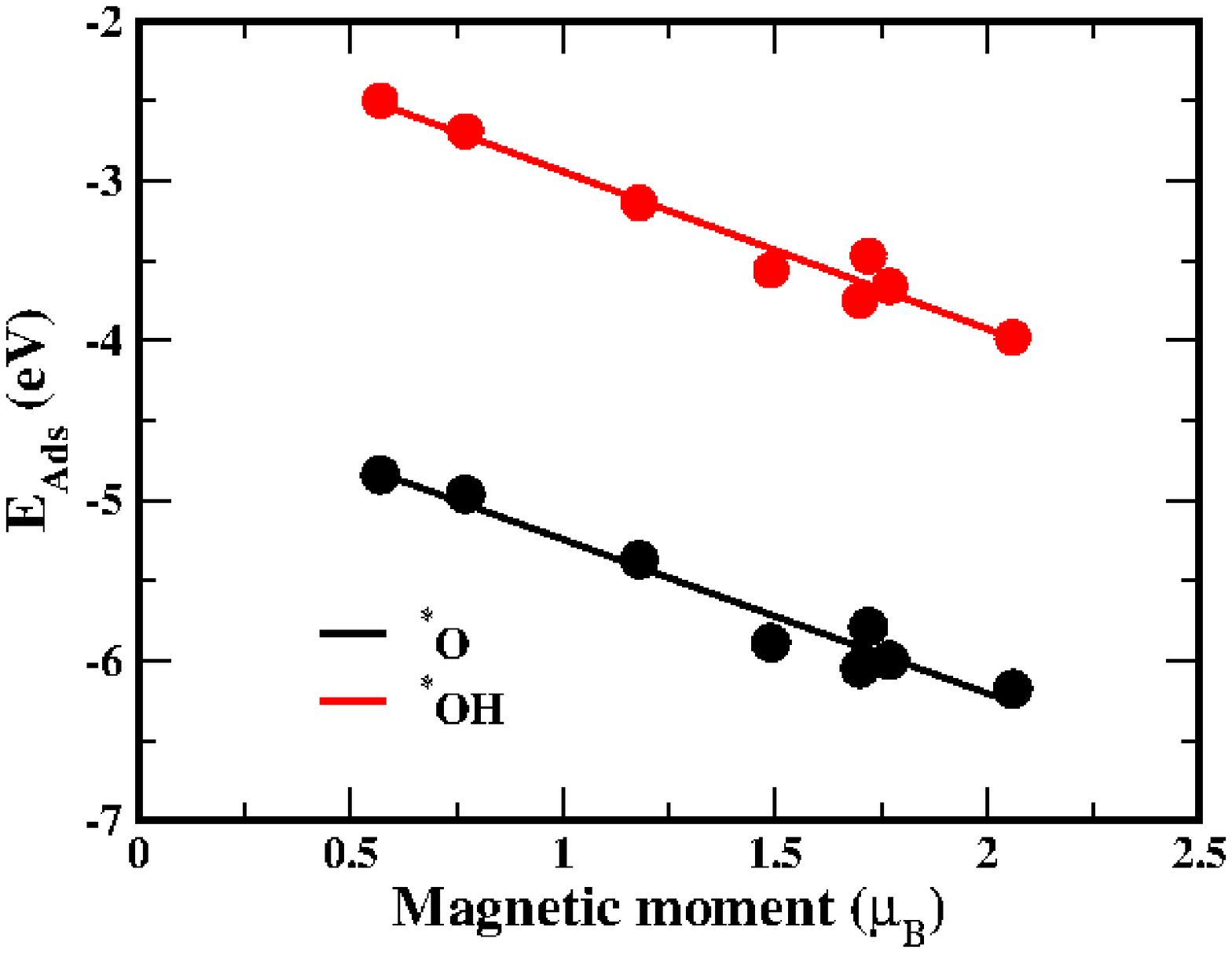}}
 \caption{Adsorption energies as a function of (a) work function of slab ($\phi$(sb)), (b) the average  valence electron ($NV_{av}$) and (c) magnetic moment of slab (m(sb)) of bimetallic magnetic TM surface. The adsorption energy scale linearly with all parameters in both O and OH adsobate. }
\end{figure}

\begin{table}[]
\caption{Correlation between selected parameters, $\lbrace\omega_i\rbrace$ and the adsorption energy of O and OH adsorbate.}
\label{tab:my-table}
\begin{tabular}{|c|ccc|lll|}
\hline
\multicolumn{1}{|c|}{\multirow{2}{*}{Plots}} & \multicolumn{3}{c|}{O}                                                                   & \multicolumn{3}{c|}{OH} \\ 
\cline{2-7} 
\multicolumn{1}{|c|}{} & \multicolumn{1}{c|}{slope} & \multicolumn{1}{c|}{intercept} & \multicolumn{1}{c|}{R$^2$} & \multicolumn{1}{l|}{slope} & \multicolumn{1}{l|}{intercept} & \multicolumn{1}{l|}{R$^2$} \\ \hline
$\phi$(sb) vs E$_{ads}$     & 1.56 & -13.34 & 0.89  & 1.6  & -11.28 & 0.87  \\
NV$_{av}$ vs E$_{ads}$        & 0.87 & -13.6  & 0.86  & 0.89 & 11.5   & 0.86  \\
m(sb) vs E$_{ads}$          & -0.96 & -4.28 & 0.95  & -0.98 & -1.96 & 0.96  \\
\hline                    
\end{tabular}
\end{table}

From the corresponding Pearson correlation data matrix, a multiple regression was carried out to determine the individual contribution towards the adsorption process. Multiple regression analysis gives estimates for the coefficients as below
\begin{subequations}

\begin{equation}
    \Delta E_{OH} = a_0 + a_1\phi (sb) + a_2N_{av} + a_3 m(sb) 
\end{equation}

\begin{equation}
\Delta E_{O} = a^{'}_0 + a^{'}_1\phi (sb) + a^{'}_2N_{av} + a^{'}_3 m(sb)
\end{equation}
\label{equ:multi}
\end{subequations}
where $a_i$ and $a^{'}_{i}$, with $i$ = 0,1,2,3, are the regression coefficients for OH and O adsorption respectively.

\section{Charge transfer through DDEC6 and Bader analysis method}
To compare our result we have reported the values calculated from both DDEC6 and Bader analysis. The trend in CT is similar in both the case, only the magnitude of CT is found to be different. 
For any electronic charge partitioning scheme, the absolute
magnitude of the charges in the particular system is of less importance than their relative
charges.  The calculated values are reported in Table-IV.

\begin{table}
 \begin{center}
 \caption{Adsorption energy $E_{ads}$, DDEC6 and Bader charges of O atom in $ ^{*} $O and $ ^{*} $OH adsorption. The charge on H atoms found to be similar ( $\sim$ 0.3 ) for all the system. }
  \begin{tabular}{|c|c|c|c|c|c|c|}
    \hline
    \multirow{2}{*}{Systems} & 
     \multicolumn{2}{c|}{E$_{ads}$ (eV)} &
      \multicolumn{2}{c|}{DDEC6}  & 
           \multicolumn{2}{c|}{Bader}  \\
      \cline{2-7}
      & $ ^{*} $O  & $ ^{*} $OH &{q(O):O} & {q(O):OH} & {q(O):O} & {q(O):OH}  \\
      \hline
    MnPd     & -6.08 & -3.98 & -0.36 & -0.56 & -0.91   & -1.18   \\
    MnPt     & -6    & -3.66 & -0.32 & -0.56 & -0.87   & -1.15 \\
    FePd     & -6.05 & -3.75 & -0.36 & -0.55 & -0.92   & -1.17   \\
    FePt     & -5.79 & -3.47 & -0.34 & -0.54 & -0.89   & -1.16   \\
    Co$_3$Pt & -5.89 & -3.56 & -0.36 & -0.56 & -0.89   & -1.19  \\
    CoPt     & -5.37 & -3.14 & -0.31 & -0.52 & -0.83   & -1.15  \\
    CoPt$_3$ & -4.96 & -2.96 & -0.27 & -0.51 & -0.75   & -1.09  \\
    NiPt     & -4.84 & -2.5  & -0.25 & -0.53 & -0.73   & -1.11  \\
       \hline
            
  \end{tabular}
 \end{center}
\end{table}

\newpage

\begin{figure}[]
  \centering
  \subfigure[]{\includegraphics[scale=0.5]{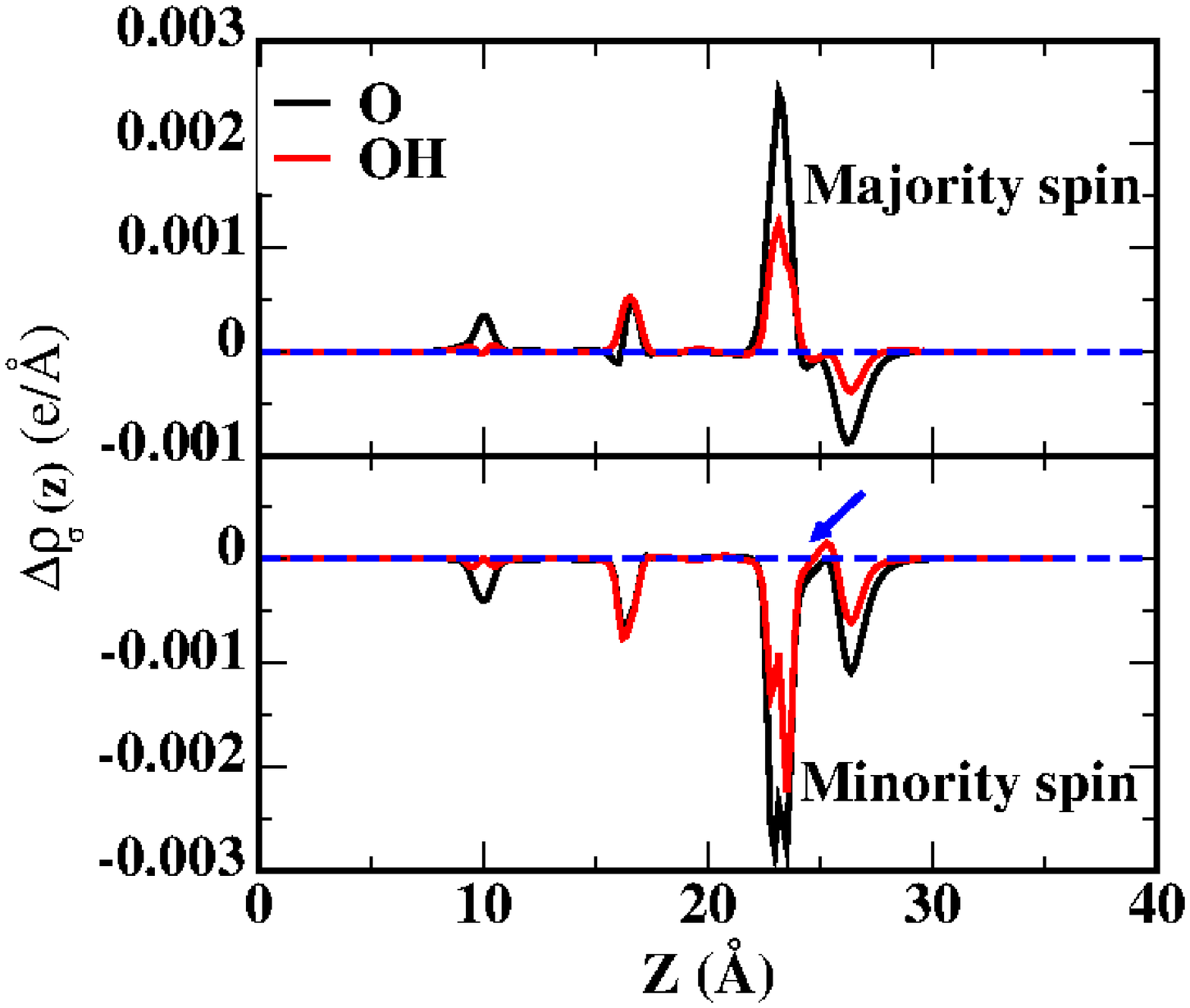}}\quad
 \caption{The line profiles of the plane-averaged charge transfer $\Delta\rho_{\sigma}(z)$ between an adatom and FePt magnetic surface as a function of the z coordinate perpendicular to the surface, , where  $ \Delta\rho_{\sigma}(z) = \int_{A_{xy}} \Delta\rho_{\sigma}(r) dx dy $. The arrow mark show the charge depletion in interface region for $ ^{*} $OH adsorption. }
 \label{CDDxy}
\end{figure}
\newpage
